\documentclass[prd, aps, superscriptaddress, preprintnumbers, twocolumn, floatfix, nofootinbib]{revtex4-1}

\usepackage{amssymb}
\usepackage{amsmath}    % need for subequations
\usepackage{graphicx}   % need for figures
\usepackage{verbatim}   % useful for program listings
\usepackage{color}      % use if color is used in text
\usepackage{subfigure}  % use for side-by-side figures
\usepackage{hyperref}   % use for hypertext links, including those to external documents and URLs
\usepackage{float}     %importanto for proper figure placement 
\usepackage{mdwlist}

\begin{document}

\title{Cosmic String Wake Detection using 3D Ridgelet Transformations}

\author{Samuel Laliberte}
\affiliation{Physics Department, McGill University, 3600 University Street, Montreal, QC, H3A 2T8, Canada}
\author{Robert H. Brandenberger}
\affiliation{Physics Department, McGill University, 3600 University Street, Montreal, QC, H3A 2T8, Canada}
\author{Disrael Camargo Neves da Cunha}
\affiliation{Physics Department, McGill University, 3600 University Street, Montreal, QC, H3A 2T8, Canada}

\begin{abstract}
Three-dimensional ridgelet statistics are used to search for the signals of cosmic string wakes
in the distribution of dark matter. We compare N-body simulations of the dark matter distribution
in cosmological models with and without a cosmic string wake, assuming that the dominant
sources of fluctuations are those predicted in the standard $\Lambda$CDM model. Cosmic
string wakes lead to overdense regions with planar topology, and hence three-dimensional
ridgelet statistics are a promising analysis tool. The string signal is easier identifiable for
larger string tensions and at higher redshift. We find that a wake produced by a string of tension 
$G\mu = 10^{-7}$ (a value slightly lower than the best current robust upper bound)
can be detected at $6 \, \sigma$ confidence level at a cosmological redshift $z = 10$.
\end{abstract}


\maketitle

\section{Introduction}

Cosmic strings \cite{Cosmic Strings} are topological defects \cite{Topological Defects} which arise in a range of relativistic quantum field theory models (for reviews see \cite{Topological Defects} and \cite{qft}) beyond the Standard Model of particle physics.  Good analogs of cosmic strings are vortex lines in superfluids and superconductors.  Like their condensed matter counterparts, cosmic strings form lines of trapped energy density.  This energy density can curve space-time and have important effects in cosmology \cite{cosmo_effect}.

Cosmic strings are relativistic objects that can be described by a unique number $\mu$.  This quantity is the mass per unit length of the string, which is also equal to its tension.  Alternatively, the string can be described by the dimensionless number $G\mu$, where $G$ is Newton's gravitational constant. The value of $\mu$ is determined by the energy scale $\eta$ at which the cosmic string is formed via the relationship \cite{Topological Defects}
\begin{equation}
\mu \sim \eta^{2} \text{.}
\end{equation}
The cosmological signatures of cosmic strings are thus more substantial for larger values of $\mu$ which implies larger values of the energy scale $\eta$.  Hence, searching for cosmic strings is a way to probe for new physics beyond the Standard Model of particle physics ``from top down'', in contrast to accelerator experiments which are more sensitive to new physics at lower energy scales.  

Cosmic strings lead to specific non-Gaussian signals in cosmic microwave background (CMB) temperature anisotropy maps, namely lines across which the temperature jumps by a value proportional to $G \mu$ \cite{KS,Moessner}. Edge detection algorithms \cite{Canny} as well as wavelet and curvelet statistics \cite{Lukas, PlanckCS} have been shown to be promising ways to search for these signals, and machine learning techniques  \cite{Oscar} have also recently been shown to have great promise. The current robust limit \footnote{There are stronger limits which come from pulsar timing surveys \cite{pulsar}, but these depend on assumptions about the distribution of string loops which are not universally accepted.} on the cosmic string tension is \cite{constraint}
\begin{equation}
G\mu \,\, \textless \,\, 1.5 \, \text{x} \, 10^{-7} \text{,}
\end{equation}
which rules out some Grand Unified particle physics models with very high scale symmetry breaking.  This limit comes from the observational upper bound on the contribution of cosmic strings to the angular power spectrum of cosmic microwave background (CMB) anisotropies obtained by combining results of the WMAP satellite \cite{WMAP} with those of the South Pole Telescope \cite{South_Pole}.  Both improving the constraint on the cosmic string tension or detecting the signature of cosmic strings would help to constrain particle physics at high energy scales.

Cosmic strings come in two different forms: loops and infinite segments \cite{Topological Defects}.  Cosmic string loops are formed when the infinite segments self-intersect.  These loops then oscillate because of their tension and slowly decay by emitting gravitational waves.  Numerical simulations lead to the conclusion that the number $N$ of long string segments that pass through any Hubble volume is of order $N \sim 10$ \cite{string_number}.   This is the so-called ``cosmic string scaling solution''. String segments which are present between the time $t_{eq}$ of equal matter and radiation and the present time $t_0$ and which our past light cone intersects produce wakes, overdense regions of dark matter and (after the time of recombination) baryons which lead to signatures in the large-scale structure of the Universe  \cite{new_obs}. Since large-scale structure observations yield three-dimensional maps (position in the sky and redshift), they potentially contain more information than the two-dimensional CMB maps. 

In this work, we will be interested in the signatures of the long cosmic string segments. Constraints on the string tension derived this way will be more robust than those which make use of assumptions about the distribution of string loops.

The signatures of cosmic string wakes are highly non-Gaussian and have specific patterns in position space. Hence, position space-based algorithms will be more effective in searching for the signals of cosmic strings than traditionally used Fourier space techniques. Another advantage of working with position space analyses is that the resulting bounds on the string tension are less sensitive to the unknown number $N$ than analyses operating in Fourier space. This is because we are looking for signals of individual strings (which are independent of $N$ modulo superposition effects) rather than for signals in correlation functions (which depend strongly on $N$). 

String wakes are nonlinear from the outset, while the fluctuations in the Standard $\Lambda$CDM model begin as Gaussian perturbations in the linear regime. On the other hand, at late times the non-linearities from the $\Lambda$CDM fluctuations become dominant \cite{Disrael2}. Hence, searching for strings in high redshift data is in principle an easier avenue. For example, string wakes lead to narrow wedges in 21 cm redshift maps 
(at redshifts larger than that of reionization) with extra absorption \cite{Holder2}. On the other hand, data is harder to obtain at high redshifts, and the measurement errors are larger. Hence, a key goal is to probe down to which redshift any given statistic is able to extract wake signals for a fixed $G\mu$. In this work, we will study the distribution of dark matter. This could in principle be measured through weak lensing surveys. If baryons follow the dark matter distribution, then we could also probe the model predictions through large-scale galaxy redshift surveys and lower redshift 21cm studies. 

At low redshifts, the density field is highly nonlinear on scales relevant to current cosmological observations of the distribution of galaxies. Hence, numerical simulations are required in order to study the predicted signals. In a recent work, a state-of-the-art N-body code \cite{CUBEP3} was extended to include the effects of a cosmic string wake \cite{Disrael}. These effects were added to the initial fluctuations from a $\Lambda$CDM cosmology. Results of runs with and without string wakes were compared, making use of a variety of specially designed statistics, and it was found that string wakes are identifiable for a string tension of $G \mu = 10^{-7}$ down to a redshift of $z = 7$. Wakes are nonlinear density perturbations present at arbitrarily early times with a distinctive geometric pattern in position space.  In particular, the planar geometry of the wake suggests that such objects could be detected using 3D ridgelet statistics.  In order to test this hypothesis, we analyzed part of the ridgelet spectrum of multiple simulated cosmic strings wakes in cosmological N-body simulations.

The conclusion of our analysis is the following.  The full 3D ridgelet transform is hard to compute as resolving a weak wake signal requires a very precise analysis.  Since the ridgelet transform has four parameters, the time required to compute the transform scales as $\mathcal{O}(n^{4})$ where $n$ is the number of values probed in our analysis for a given parameter.  This complicates the analysis on standard computers.  However, a partial ridgelet transform analysis shows that a cosmic string of tension $G\mu = 10^{-7}$ can be detected at a high significance ($5 \,\sigma$ level) at a redshift of $ z = 10$. This bound is competitive to what was obtained in \cite{Disrael}. 

The outline of this paper is as follows:  In Sections 2 and 3, we present a brief review of cosmic string wakes and how they can be recreated in cosmological N-body simulations.  Then, we discuss the 3D ridgelet transform and its implementation in cosmological N-body simulation in Section 4.  Finally, we show how cosmic string wakes appear in ridgelet space in Section 5 and present our results for the detection of weak wake signals in Section 6. We use units of which the speed of light $c$ is set to $c = 1$. We assume a homogeneous and isotropic cosmological background with vanishing spatial curvature and scale factor $a(t)$, where $t$ is physical time. We set the scale factor to be $a(t_0) = 1$ at the present time $t_0$. Thus, comoving lengths correspond to physical lengths today. The Hubble expansion rate is taken to be $h \times 100{\rm{km}} \, s^{-1} {\rm{Mpc}}^{-1}$, where $h$ is a constant.

%%%%%%%%%%%%%%%%%%%%%%%%%%%%%%%%%%
%% Cosmic String Wake Formation %%
%%%%%%%%%%%%%%%%%%%%%%%%%%%%%%%%%%

\section{Cosmic String Wake Formation}

\begin{figure}[h]
	\centering
	\includegraphics[width=8cm]{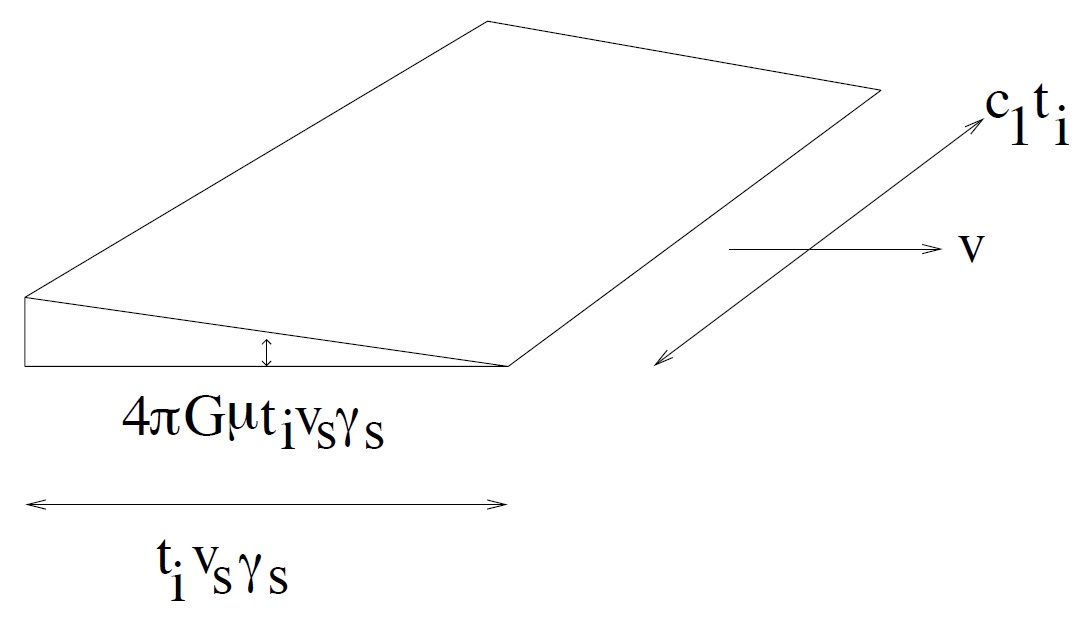}
	\caption{Geometry of a cosmic string wake \cite{wakes}. In the figure, $v_{s}$ is the string velocity $v$, $\gamma_{s}$ is the gamma factor $\gamma(v)$ and $c_{1}$ is a constant of order unity that depends on the length of the string.}
	\label{fig:wakes}
\end{figure}

Space perpendicular to a long straight cosmic string segment is conical with a ``deficit angle'' \cite{deficit} given by
\begin{equation}
\alpha \, = \, 8 \pi G \mu \, .
\end{equation}
For strings forming in a phase transition, this conical structure extends to a Hubble length from the string \cite{Joao}. Hence, when a long string segment moves through a uniform matter distribution of the early universe, the matter behind the string acquires a velocity perturbation
\begin{equation}
\delta v \, = \, 4 \pi v\gamma(v) G\mu
\label{vpert}
\end{equation}
towards the plane spanned by the tangent vector to the moving string and the direction of motion, where $v$ is the velocity of the string and $\gamma(v) = 1/\sqrt{1-v^{2}}$.  This, in turn, leads to a wedge-shaped overdensity (density being twice the background density) behind the string, a ``wake'' \cite{wakes}.  A cosmic string at the time $t_{i}$ will lead to a wake with comoving size
\begin{equation}
c_{1}t_{i} \, \text{x} \, v\gamma(v)t_{i} \, \text{x} \, \delta v t_{i} \, \text{,}
\end{equation}
where the factors are, from left to right, the length, the depth and the mean width of the wake.  Here, $c_{1}$ is a constant of order 1.  The geometry of the wake is shown in Figure \ref{fig:wakes}.

Once formed, the cosmic string wake grows by gravitational accretion, which is studied using the Zel'dovich approximation \cite{Zeldovich}.  The idea behind this approximation is to consider a thin shell of matter which is located initially at a physical height 
\begin{equation}
H(t_{i}) \, = \, a(t_{i})q 
\end{equation}
above the center of the wake, where $t_{i}$ is the time when the wake is laid down.  Here, $a(t_{i})$ is the cosmological scale factor evaluated at time $t_{i}$ and $q$ is the initial comoving height.  As a consequence of the gravitational pull of the matter overdensity inside the wake, a comoving displacement $\psi(t)$ gradually builds up (where $\psi(t_{i})=0$).  The physical height at time $t \, \textgreater \, t_{i}$ then can be written as 
\begin{equation}
H(q,t) \, = \, a(t)(q - \beta) \text{,}
\end{equation}
where $\beta$ is a comoving displacement.  If matter accretes via Newtonian gravity, the height of the wake at a later time is determined by
\begin{equation}
\ddot{H} \, = \, - \frac{\partial \Phi}{\partial H} \text{,}
\end{equation}
where $\Phi$ is the Newtonian gravitational potential which is determined by the Poisson equation in terms of the mass overdensity.  We then calculate the value $q(t)$ (which we call $q_{nl}(t,t_{i})$) for which the shell stops growing in size at time $t$. This is given by
\begin{equation}
\dot{H}(q(t),t) = 0 \, \text{.}
\end{equation}
After this turnaround point, the shell virializes at a physical height which is half the value at its maximum.  This virialized region forms the wake.  For a cosmic string forming a perfectly straight line, the wake will take the form of a region of planar overdensity.  This region of planar overdensity as a comoving height that grows linearly in the scale factor.  As a result of a straight-forward computation, we obtain 
\begin{equation}
q_{nl}(t,t_{i}) \, = \, \frac{a(t)}{a(t_{i})}\frac{24\pi}{5}v\gamma(v)G\mu(z(t_{i})+1)^{-1/2}t_{i} \text{,}
\end{equation}
which gives half the height the shell would have if it was simply expanding with the Hubble flow.  Note that $z(t)$ is the cosmological redshift.  Finally, the comoving planar dimension of the wake formed at time $t_i$ is given by the comoving horizon at $t_{i}$, namely
\begin{equation}
d \, = \, z(t_{i})^{-1/2}t_{0} \, \text{.}
\label{comdist}
\end{equation}

A string segment only lives for one Hubble expansion time (before a string intersection occurs). However, since cosmic string wakes are made of accreted matter, they persist after the string segment has decayed. String wakes whose world sheet intersects the past light cone lead to an observable signal. The following section will explain how to recreate this signal in cosmological N-body simulations.

%% SL: I changed the variable for the physical height from h to H to avoid confusion with the Hubble parameter h introduced in the next section.

%%%%%%%%%%%%%%%%%%%%%%%%%%%%%%%%%%%%%%%%%%%%%%%
%% Cosmic String Wakes in N-body simulations %%
%%%%%%%%%%%%%%%%%%%%%%%%%%%%%%%%%%%%%%%%%%%%%%%

\section{The Cosmological N-body simulations} \label{sec:sim}

The cosmological N-body simulations consist of a three-dimensional cubic box with a set of N points (representing equal mass particles) labeled by an index $i$ represented by their coordinates $\vec{x}_{i}$ and velocities $\vec{v}_{i}$.  After setting up the initial distribution of points at redshift $z_I$ (which is in general smaller than the redshift $z(t_i)$ when the wake is assumed to have been created),  they are evolved using the Newtonian gravitational force equations to a later time $t_w$ when a cosmic string wake is inserted.  Once the wake is inserted, we once more let the points in the box evolve according to Newtonian gravity to our current time while keeping track of the position and the velocity of the particles inside the box throughout the process.  The simulations are described in more detail in \cite{Disrael}.  In this section, we present information about the data boxes relevant to the analyses presented in Sections 5 and 6. 

The simulations are produced with a public high-performance cosmological N-body code named CUBEP$^{3}$M \cite{CUBEP3}.  A data box is segmented in multiple cells and the number of particles per dimension introduced inside the box is chosen to be half the number of cells per dimension.  To place the particles in the box, the initial conditions generator of the program reads a transfer function constructed with the CAMB online toolkit\footnote{CAMB:https://lambda.gsfc.nasa.gov/toolbox/tb camb form.cfm} and lays out a distribution of points corresponding to $\Lambda$CDM fluctuations \cite{CDM_fluc} at the initial redshift $z_{I}$ with the following cosmological parameters: $\Omega_{\Lambda} = 0.7095$, $\Omega_{b} = 0.0445$, $\Omega_{CDM} = 0.246$, $n_{t} = 1$, $n_{s} = 0.96$, $\sigma_{8} = 0.8628$, $h = 0.70$ and $T_{CMB}(t_{0}) = 2.7255$.  Here, $\Omega_{\Lambda}$ is the energy fraction of dark energy, $\Omega_{b}$ is the energy fraction of baryonic matter, $\Omega_{CDM}$ is the energy fraction of cold dark matter, $n_{t}$ is the tensor spectral index, $n_{s}$ is the scalar spectral index, $\sigma_{8}$ is the amplitude of the linear power spectrum on the scale of 8 $h^{-1} {\rm{Mpc}}$, h is the Hubble parameter and $T_{CMB}(t_{0})$ is the temperature of the cosmic microwave background at our current time $t_{0}$.  Each particle inside the box has a distinct ID number associated to it which allows us to track its position inside the box.  The initial redshift $z_I$ is chosen at a point in time when the density fluctuations are in the linear regime.

As mentioned before, the particles move according to the gravitational interaction between them throughout the simulation.  (See \cite{CUBEP3} for more information on how the gravitational attraction on each particle is computed.)  The wake is introduced at a later time $t_w$ after $t_I$.  To produce a particle distribution corresponding to a wake overdensity, the particles are moved and given a velocity kick towards the central plane $y = 0$ $h^{-1}$Mpc in the simulation box.  The goal of this process is to simulate the velocity perturbation $\delta v$ given by equation \ref{vpert}.  We consider wakes laid down at the time of equal matter and radiation $t_{eq}$ because they have had more time to grow in thickness than those created later, and since they are the largest among those present at $t_{eq}$. Since the comoving planar distance of such a wake, given by equation \ref{comdist}, is much bigger than the size of the simulation box, it is justified to insert the velocity perturbation as a planar perturbation.  The exact magnitude of the velocities and displacements given to the particles are calculated according to the Zel'dovich approximation mentioned in the previous section, evolving the fluctuation from $t_i = t_{eq}$ to the time $t_w$ of wake insertion.  As a result of this computation, the comoving displacement $\psi(t)$ of the particles towards the plane at times $t \,\textgreater \, t_{i}$ is given by
\begin{equation}
\psi(t) \, = \,  \frac{3}{5}4\pi G\mu v \gamma(v)t_{i}z(t_{i})\frac{z(t_{i})}{z(t)} \, \text{.}
\label{displacement}
\end{equation}
The last factor represents the linear theory growth of the fluctuation while the other factor of $z(t_{i})$ represent the conversion from physical to comoving velocity.  The comoving velocity perturbation is
\begin{equation}
\dot{\psi}(t) \, = \, \frac{2}{5}4\pi G\mu v \gamma(v)t_{i}z(t_{i})\frac{z(t_{i})}{z(t)}\frac{1}{t} \, \text{.}
\label{velovity_kick}
\end{equation}
In the context of the simulations, the displacement and the velocity perturbation given to the particles towards the central plane $y = 0$ $h^{-1} {\rm{Mpc}}$ are computed respectively from equation \ref{displacement} and \ref{velovity_kick} at the time $t=t_w$ when the wake is inserted.

After the wake insertion, the modified data cube is evolved using CUBEP$^{3}$M until redshift $z=0$.  This way, the behavior of the wake overdensity can be studied at lower redshift until it is completely disrupted by the other density fluctuations \cite{Disrael2}.  The next section will introduce the statistics used in order to study the wake overdensity.

%%RB: need to mention the box size => SL: The box size changes depending on the simulation.  I introduce it at the begining of each analysis.
%%RB: check for text overlap in the beginning part of this section => SL: What kind of overlap?

%%%%%%%%%%%%%%%%%%%%%%%%%%%%%%%%
%% The 3D Ridgelet Transform  %%
%%%%%%%%%%%%%%%%%%%%%%%%%%%%%%%%

\section{The 3D Ridgelet Transform}

Like the Fourier transform, which is an orthogonal projection of a function onto the space of phasors $e^{ikx}$, the 3D ridgelet transform \cite{3D Sparse Representations} is an orthogonal projection of a function on the space of ridgelet functions $\Psi(\vec{x})$, which are wavelet functions $\psi$ constant along a plane with normal vector
\begin{equation}
\vec{n}(\theta_{1},\theta_{2}) \, = \, (\cos \theta_{1} \sin \theta_{2}, \sin \theta_{1} \sin \theta_{2}, \cos \theta_{2}) \, \text{.}
\end{equation}
Here, the angles $\theta_{1} \in [0,2\pi[$ and $\theta_{2} \in [0,\pi[$ determine the orientation of the plane in spherical coordinates.  For each plane with normal vector $\vec{n}(\theta_{1},\theta_{2})$, we can define the trivariate ridgelet function evaluated at $\vec{x}$ by
\begin{equation}
\Psi_{a,b,\theta_{1},\theta_{2}}(\vec{x}) \, = \, a^{-1/2}\psi\left(\frac{\vec{x}\cdot\vec{n}(\theta_{1},\theta_{2})-b}{a}\right) \, \text{,}
\end{equation}
where $a$, which satisfies $a \, \textgreater \, 0$, is a scale parameter and $b \in {\Bbb R}$ determines the position of the ridgelet function.  Given an integrable trivariate function $f(\vec{x})$, its 3D ridgelet coefficients are defined by:
\begin{equation}
\mathcal{R}_{f}(a,b,\theta_{1},\theta_{2}) \, = \, \int_{\mathbb{R}^{3}}^{} f(\vec{x}) \Psi_{a,b,\theta_{1},\theta_{2}}(\vec{x})d\vec{x} \text{,}
\label{eq:coef}
\end{equation}
Computing the ridgelet transform of a function means computing the ridgelet coefficients $\mathcal{R}_{f}$ for all possible values of $a$, $b$, $\theta_{1}$ and $\theta_{2}$, which constitutes the ridgelet space.

Given the planar geometry of the ridgelet function, the 3D ridgelet analysis can be constructed as a wavelet analysis in the Radon domain.  In 3D, the Radon transform $\textbf{R}(f)$ of $f$ is the collection of hyperplane integrals indexed by the orientation $(\theta_{1},\theta_{2})$ in spherical coordinates and a position coefficient $t \in {\Bbb R}$.  The value of $\textbf{R}(f)$ is given by
\begin{equation}
\textbf{R}(f)(\theta_{1},\theta_{2},t) \, = \,  \int_{{\Bbb R}^{3}}^{}f(\vec{x})\delta(\vec{x}\cdot\vec{n}(\theta_{1},\theta_{2})-t)d\vec{x} \text{,}
\end{equation}
where $\delta$ is the Dirac delta function.  Then, the 3D ridgelet transform is exactly the application of a 1D wavelet transform along the slices of the Radon transform where the orientation $(\theta_{1},\theta_{2})$ is kept constant but $t$ is varying:
\begin{equation}
\mathcal{R}_{f}(a,b,\theta_{1},\theta_{2}) \, = \, \int\psi_{a,b}(t)\textbf{R}(f)(\theta_{1},\theta_{2},t)dt \text{,}
\label{coeff}
\end{equation}
where $\psi_{a,b} = \psi((t-b)/a)/\sqrt{a}$ is a 1-dimensional wavelet.  Therefore, a good strategy for calculating the continuous ridgelet transform in 3D is to compute the Radon transform $\textbf{R}(f)$ first and then apply a 1-dimensional wavelet to the slices defined by fixing the orientation $(\theta_{1},\theta_{2})$ in $\textbf{R}(f)$.

In order to define a way to perform a ridgelet transformation on the 3D data box which constitutes the cosmological N-body simulation, each point $i$ at a position $\vec{x_{i}}$ inside the box can be considered locally as a Dirac delta function in the energy density $\rho$.  Using this representation, the local density $\rho(\vec{x})$ inside the box is given by
\begin{equation}
\rho(\vec{x}) \, = \, \sum_{i=1}^{N}\delta(\vec{x}-\vec{x_{i}}) \, \text{.}
\end{equation}
Here, the local density is normalized in a way that the total mass $M$ of the particles in the box is a dimentionles quantity equal to the number $N$ of particles in the box.  This way, each particle has a mass $m = 1$.  To perform the Radon tranform at a specific orientation $(\theta_{1},\theta_{2})$ inside the box, it suffices to redefine the position of each point inside the box as their orthogonal projection $t_{i}(\theta_{1},\theta_{2}) = \vec{x_{i}}\cdot\vec{n}(\theta_{1},\theta_{2})$ on a line spanned by $\vec{n}(\theta_{1},\theta_{2})$ passing through the center of the box.  We obtain
\begin{equation}
\textbf{R}(\rho)(\theta_{1},\theta_{2}) \, = \,  \sum_{i=1}^{N}\delta(t-t_{i}(\theta_{1},\theta_{2})) \, \text{.}
\end{equation}
Finally, each ridgelet coefficient can be computed using equation \ref{coeff}.  The expression for the coefficients trivially reduces to
\begin{equation}
\mathcal{R}_{\rho}(a,b,\theta_{1},\theta_{2}) \, = \, \sum_{i=1}^{N}\psi_{a,b}(t_{i}(\theta_{1},\theta_{2})) \, \text{.}
\end{equation}

\begin{figure}[!h]
	\centering
	\includegraphics[width=6.2cm]{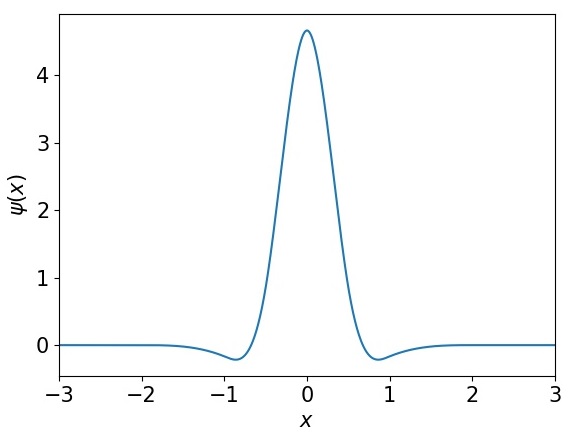}
	\caption{Plot of the wavelet function $\psi(x)$ in dimensionless units.}
	\label{fig:wavelet}
\end{figure}

A good choice of wavelet function $\psi$ in the expression of $\psi_{a,b}$ is one that satisfies the following difference between two scaling functions $\phi$:
\begin{equation}
\frac{1}{8}\psi\left(\frac{x}{2}\right) \, = \, \phi(x) - \frac{1}{8}\phi\left(\frac{x}{2}\right) \text{.}
\end{equation}
Here, the chosen scaling function $\phi$ is a B-spline of order 3:
\begin{equation}
\begin{split}
\phi(x) &= \frac{1}{12}(|x-2|^{3}-4|x-1|^{3}+6|x|^{3} \\
        &-4|x+1|^{3}+|x+2|^{3}) \, \text{.}
\end{split}
\end{equation}
The B-spline, defined by the equation above, is shown in Figure \ref{fig:wavelet}.

Each ridglet coefficient has 4 independent parameters $a \, \textgreater \, 0$, $b \in {\Bbb R}$, $\theta_{1} \in [0,2\pi[$ and $\theta_{2} \in [0,\pi[$. If we discretize ridgelet space into $n$ intervals along each axis, then the time required to perform the ridgelet tranformation grows as $\mathcal{O}(n^{4})$.  This also means that the time required to perform the ridgelet transformation grows as $\mathcal{O}(\text{N}^{4})$ for a fixed value of $n$ and a varying number of particles N.  Therefore, the computation is very expensive for a high number of particles. 

In order to analyze the cosmic string wake signature in ridgelet space, we will for the sake of simplicity restrict the numbers of unknown arguments in the ridgelet transformation This will yield a bound on the string wake detection efficiency, as will be discussed in the following section.

%%%%%%%%%%%%%%%%%%%%%%%%%&%%%%%%%%%%%%
%% Wake Signature in Ridgelet Space %%
%%%%%%%%%%%%%%%%%%%%%%%%%%%%%%%%%%%%%%

\section{Plane Wakes signature in partial ridgelet transformations}

The main idea behind using the 3D transform in order to detect cosmic string wakes is to find a subspace of the ridgelet space where the wake appears as a maximum in the ridgelet coefficients $\mathcal{R}_{\rho}$.  A good way to do this is to perform the ridgelet transformation for a fixed value of the parameters $a$ and $b$ while varying the orientation $(\theta_{1},\theta_{2})$.

The values of $a$ and $b$ are chosen in a way that maximizes the ridgelet coefficients.  Since the wake and the ridgelet function $\Psi$ have planar geometry, we expect that the ridgelet coefficients $\mathcal{R}_{\rho}$, which are an inner product of the density $\rho$ with $\Psi$, will be optimized for the parameters $a$, $b$, $\theta_{1}$ and $\theta_{2}$ that match the characteristic width and position of the wake.  That is, we expect that a value of $a$ close to the width of the wake, a value of $b$ close to the position of the wake and an orientation $(\theta_{1},\theta_{2})$ normal to the plane made by the wake will yield a maximum ridgelet coefficient.  In our simulation, where the center of the box is at the origin and the wake is located on the z-x plane, this means that $b = 0$ $h^{-1} {\rm{Mpc}}$, $\theta_{1} = \pi/2$, $\theta_{2} = \pi/2$ and a value of $a$ corresponding to the width of the wake should optimize the value of $\mathcal{R}_{\rho}$.

\begin{figure}[h]
	\centering
	\includegraphics[width = 8.0cm]{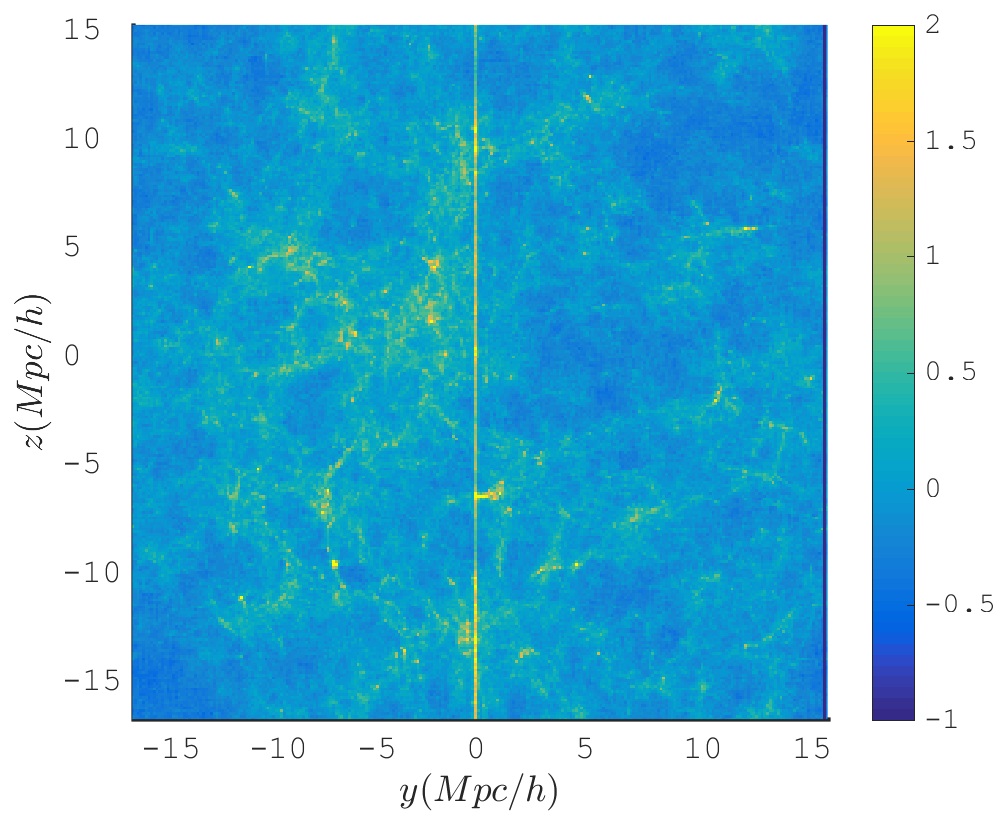}
	\caption{Density contrast of a 2D projection of the box on the y-z plane \cite{Disrael}.  The $32 \, h^{-1} {\rm{Mpc}} \, \times \, 32 \, h^{-1} {\rm{Mpc}} \, \times \,  32 \, h^{-1} {\rm{Mpc}}$ data box contains a cosmic string wake caused by a cosmic string of tension $G\mu = 10^{-6}$ at the position y = 0 $h^{-1} {\rm{Mpc}}$ and the redshift $z = 7$.  The color scheme, shown on the right, depicts the range of possible values for the fluctuation ratio $\delta S/S$.  Here, $S$ is the mean surface density and $\delta S$ is the difference between the local surface density and the mean surface density.}
	\label{fig:wake_signal1}
\end{figure}

We tested this hypothesis using a data box which describes a $32 \, h^{-1} {\rm{Mpc}} \, \times \, 32 \, h^{-1} {\rm{Mpc}} \, \times \,  32 \, h^{-1} {\rm{Mpc}}$ cubic volume.  The number of cells per dimension for the simulation and the number of particles per dimension in the box was respectively 512 and 256.  The simulations started at redshift $z_I = 63$ and the wake produced by a cosmic string of string tension $G\mu = 10^{-6}$ was introduced at redshift $z = 7$. The 2D density contrast of the box on the y-z plane at this redshift is shown in Figure \ref{fig:wake_signal1}.  
As we can see, the line overdensity visible at the position $y = 0$ $h^{-1} {\rm{Mpc}}$ is in the non-linear regime which indicates the presence of the cosmic string wake.  

\begin{figure}[h]
	\centering
	\includegraphics[width = 8.0cm]{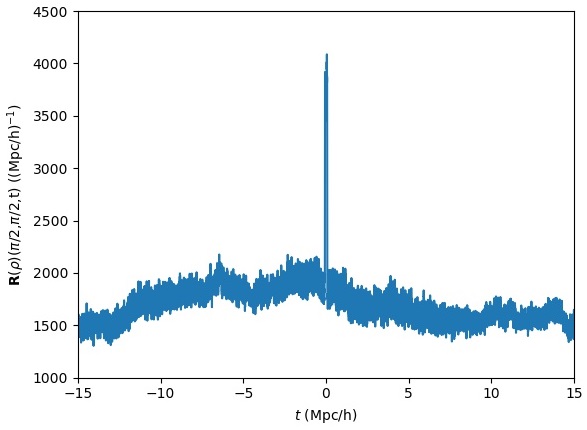}
	\caption{For a wake caused by a cosmic string of string tension $G\mu = 10^{-6}$ at redshift $z = 7$, the Radon transform $\textbf{R}(\rho) = \textbf{R}(\rho)(\theta_{1},\theta_{2},t)$ where $\theta_{1} = \theta_{2} =\pi/2$, is ploted for 10000 values of $t$ between $-15 \, h^{-1} {\rm{Mpc}}$ and $15 \, h^{-1} {\rm{Mpc}}$. For this specific orientation and position, $t$ corresponds to the $y$ axis which gives us a density projection on the y axis.  The width of the wake is $(1.1382 \, \pm \, 0.0003) \, \times \, 10^{-1} h^{-1} {\rm{Mpc}}$, where the uncertainty is defined as half the distance between two evaluated values of $t$.}
	\label{fig:radon}
\end{figure}

\begin{figure}[!h]
	\centering
	\includegraphics[width = 6.9cm]{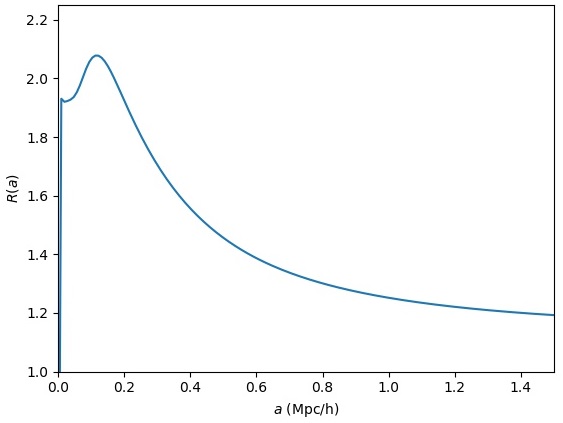}
	\caption{For a wake caused by a cosmic string of tension $G\mu = 10^{-6}$ at redshift $z = 7$, the ratio $R(a)$ is depicted for 1000 values of $a$ between 0 $h^{-1} {\rm{Mpc}}$ and 1.5 $h^{-1} {\rm{Mpc}}$.  The ratio has a maximum at $a_{max} = (1.168 \pm 0.006) \, \times \, 10^{-1} h^{-1} {\rm{Mpc}}$, where the uncertainty is defined as half the distance between two evaluated values of $a$.  As $a$ becomes larger, $R(a)$ eventually converges to 1 as the wake signal becomes indistinguishable from the Gaussian fluctuations.}
	\label{fig:scale_ratios}
\end{figure}

The analysis was performed at the redshift of insertion of the wake ($z = 7$).  Since the wake can move to another position at later redshift, this ensures that the wake is at the center of the box a the moment of the analysis.  This position can be seen in the density projection of the box on the y-axis (see Figure \ref{fig:radon}), where the peak at the origin exposes the overdensity created by the wake.
To find the value of $a$ that maximises the ridgelet coefficients, we imposed the parameters $b=0$ $h^{-1}$Mpc and $\theta_{1} = \theta_{2} =\pi/2$ and studied the behavior of $\mathcal{R}_{\rho}$ as the value of $a$ was varied.  A good way to do this is to plot the ratio of $\mathcal{R}_{\rho}$ at a specific scale $a$ with respect to the average of $\mathcal{R}_{\rho}$ over all orientations.  Assuming the ridgelet coefficients are described by $\mathcal{R}_{\rho} = \mathcal{R}_{\rho}(a,b,\theta_{1},\theta_{2})$, this ratio can be defined as
\begin{equation}
R(a) \, = \, \frac{\mathcal{R}_{\rho}(a,0,\pi/2,\pi/2)}{\langle\mathcal{R}_{\rho}(a,0,\theta_{1},\theta_{2})\rangle} \, \text{,}
\end{equation}
where $\langle\mathcal{R}_{\rho}(a,0,\theta_{1},\theta_{2})\rangle$ is the mean value of $\mathcal{R}_{\rho}$ with respect to the orientation $(\theta_{1},\theta_{2})$ and a fixed value of $a$.  For the present data box, the plot of this ratio is shown in Figure \ref{fig:scale_ratios}.
As we can see, $R(a)$ as a maximum at $a_{max} = (1.168 \pm 0.006) \, \times \, 10^{-1} h^{-1} {\rm{Mpc}}$.  This value of $a_{max}$ is close to the physical width of the wake.  Therefore, it makes sense to fix $a$ to the value of $a_{max}$ for the partial ridgelet transformation.

\begin{figure}[h]
	\centering
	\includegraphics[width = 8.0cm]{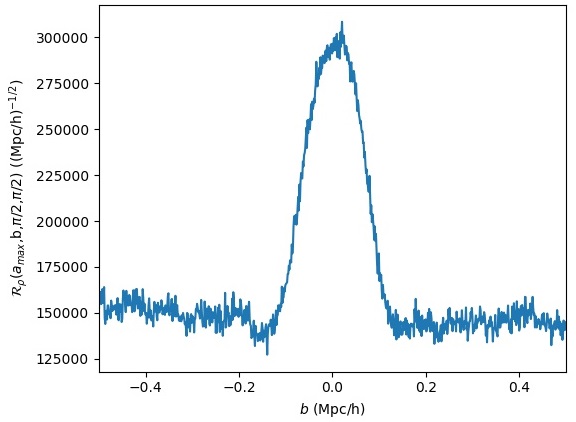}
	\caption{For a wake cased by a cosmic string of tension $G\mu = 10^{-6}$ at redshift $z = 7$, the ridgelet coefficents $\mathcal{R}_{\rho}$ are plotted for 1000 values of $b$ between $-0.5 \, h^{-1} {\rm{Mpc}}$ and $0.5 \, h^{-1} {\rm{Mpc}}$ while imposing that $a = a_{max}$, $\theta_{1} = \theta_{2} =\pi/2$.  The ridgelet coefficients have a maximum at $b_{max} = (0 \pm 8) \, \times \, 10^{-1}$ $h^{-1} {\rm{Mpc}}$.  The uncertainty is defined as half the width at half maximum of the peak.}
	\label{fig:bplot}
\end{figure}

\begin{figure}[h]
	\centering
	\includegraphics[width = 8.0cm]{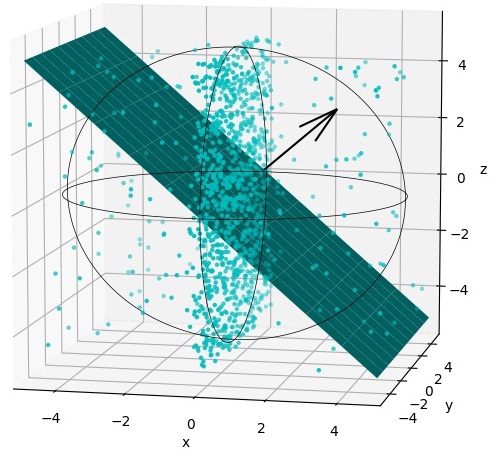}
	\caption{Sketch of the partial wavelet transformation.  The ridgelet transform is computed inside a sphere the size of the box for different orientations $(\theta_{1}, \theta_{2}) \in [0,\pi[ \, \text{x} \,  [0,\pi[$. The ridgelet function is constant along the planes normal to the direction vector $\vec{n}(\theta_{1}, \theta_{2})$ shown in black.  Therefore, we expect the ridgelet coefficients to be maximised when $\theta_{1} = \theta_{2} =\pi/2$; that is, when the plane with normal $\vec{n}(\theta_{1}, \theta_{2})$ coincides with the plane formed by the wake.}
	\label{fig:algorithm_picture}
\end{figure}

At this point, there should be no doubt that $b = 0$ $h^{-1} {\rm{Mpc}}$ is also a good parameter to fix.  However, to ensure that $b = 0$ $h^{-1} {\rm{Mpc}}$ yields a maximum coefficient in the ridgelet transformation, we plotted $\mathcal{R}_{\rho}$ for $a = a_{max}$ and $\theta_{1} = \theta_{2} =\pi/2$ and varied $b$ around the value of zero.  As shown in Figure \ref{fig:bplot}, the ridgelet coefficients have a maximum at $b_{max} = (0 \pm 8) \, \times \, 10^{-1} h^{-1} {\rm{Mpc}}$.
Since the uncertainty of $b_{max}$ includes the value of zero, $b = 0$ $h^{-1} {\rm{Mpc}}$ is a good fixed parameter.  Finally, choosing $a = a_{max}$ and $b = 0$ $h^{-1} {\rm{Mpc}}$ as fixed parameters, we can compute the partial ridgelet transformation.  To avoid any edge effects that could be caused by the geometry of the box, we only consider the points inside the largest possible sphere centered on the wake inside the box.  Then, we compute $\mathcal{R}_{\rho}$ for $n^{2}$ orientations $(\theta_{1},\theta_{2}) \in [0,\pi[ \, \times \,  [0,\pi[$.  This process is shown in Figure \ref{fig:algorithm_picture}.
Here, the orientations are constrained to a hemisphere in order to avoid any unwanted periodicities in the ridgelet coefficients.  The resulting subspace of the ridgelet space can be visualized as a surface plot. 
\begin{figure}[h]
	\centering
	\includegraphics[width = 8.0cm]{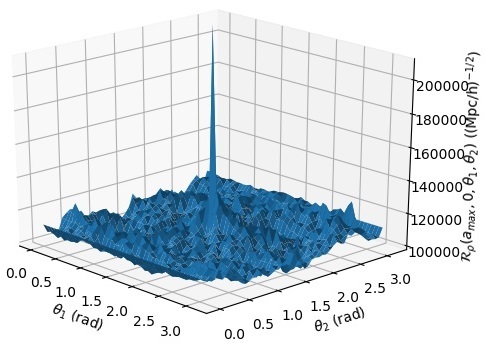}
	\caption{Surface plot of $\mathcal{R}_{\rho}$ for $a  = a_{max}$, $b = 0$ $h^{-1} {\rm{Mpc}}$ and $50^{2}$ orientations $(\theta_{1}, \theta_{2}) \in [0,\pi[ \, \times \,  [0,\pi[$.  The plot has a clear maximum at $\theta_{1} = \theta_{2} =\pi/2$ for a wake cased by a cosmic string of string tension $G\mu = 10^{-6}$ at redshift $z = 7$.}
	\label{fig:surfaceplot1}
\end{figure}
As shown in Figure \ref{fig:surfaceplot1}, the ridgelet coefficients have a clear maximum at $\theta_{1} = \theta_{2} =\pi/2$ which confirms our hypothesis that the parameters $a = a_{max}$, $b = 0$ $h^{-1} {\rm{Mpc}}$ and $\theta_{1} = \theta_{2} =\pi/2$ maximise the ridgelet coefficients.  Coincidentally, we can use this signal as a way to quantify at which level a cosmic string wake can be detected using ridgelet transformations.  In order to do this, we compute the how much the maximum value of the ridgelet coefficients varies from its mean value, then compare this observable to the same maximum value in simulations where no wake is inserted.  The steps are the following.  Define the maximum fluctuation by
\begin{equation}
\delta \mathcal{R}_{\rho_{max}} \, = \,  \mathcal{R}_{\rho_{max}} - \langle \mathcal{R}_{\rho} \rangle \text{,}
\end{equation}
where $\mathcal{R}_{\rho_{max}}$ is the absolute maximum and $\langle \mathcal{R}_{\rho} \rangle$ is the mean value of the ridgelet coefficients over all orientations in the subspace that we have defined.  Using this definition, we compute the maximum fluctuation for a simulation with a wake and multiple simulations without a wake for comparison.  Then, a good measure of detection for the cosmic string wake is the confidence level $\mathcal{C}$ which we define as
\begin{equation}
\mathcal{C} \, = \,  \frac{s-\mu}{\sigma} \, \text{.}
\end{equation}
Here, $s$ is the maximum fluctuation for a simulation with a wake and $\mu$ and $\sigma$ are respectively the mean and the standard deviation of the maximum fluctuations without a wake.  This confidence level is the number of $\sigma$'s away from the mean of maximum fluctuations without wakes.
 
%%%%%%%%%%%%%%%%%%%%%%%%%%%%%%%%%%%%%%%
%% Lower bound on the string tension %%
%%%%%%%%%%%%%%%%%%%%%%%%%%%%%%%%%%%%%%%

\section{Weak plane wakes detection in partial ridgelet transformations}

The method described in the previous section provides a way to detect wake signals which would not be visible by eye in the three-dimensional density maps.  As discussed at the beginning of the paper, it is of interest to determine the lowest value of the string tension which can be detected in data at a particular redshift or to ask down to which redshift the string wake remains visible for a fixed value of the string tension. It is this second question which we study here. We fix the string tension to be $G\mu = 10^{-7}$, a value close to but below the current limit. We find that given the limited resolution of the simulations of the present study, string wakes remain identifiable down to a redshift of $z  = 10$. We expect (as in the work of \cite{Disrael}) that with an improved resolution the string wake will remain visible to a lower redshift.  

We studied two sets of 4 data boxes which describe a $64 \, h^{-1} {\rm{Mpc}} \, \times \, 64 \, h^{-1} {\rm{Mpc}} \, \times \,  64 \, h^{-1} {\rm{Mpc}}$ cubic volume.  For both sets of simulations, the number of cells per dimensions for the simulations and the number of particles per dimension in the box was, respectively, 512 and 256, and the simulations started at redshift $z_I = 63$.  In the first set of simulations, the perturbations from a wake produced by a cosmic string of string tension $G\mu = 10^{-7}$ was introduced at redshift $z = 31$.  In the second set of simulations, no wake was introduced in the data box.  The purpose of having a second set of simulations without wake was to expose and quantify the signal difference between a simulation with a wake and without a wake.

\begin{figure}[h]
	\centering
	\includegraphics[width = 8.0cm]{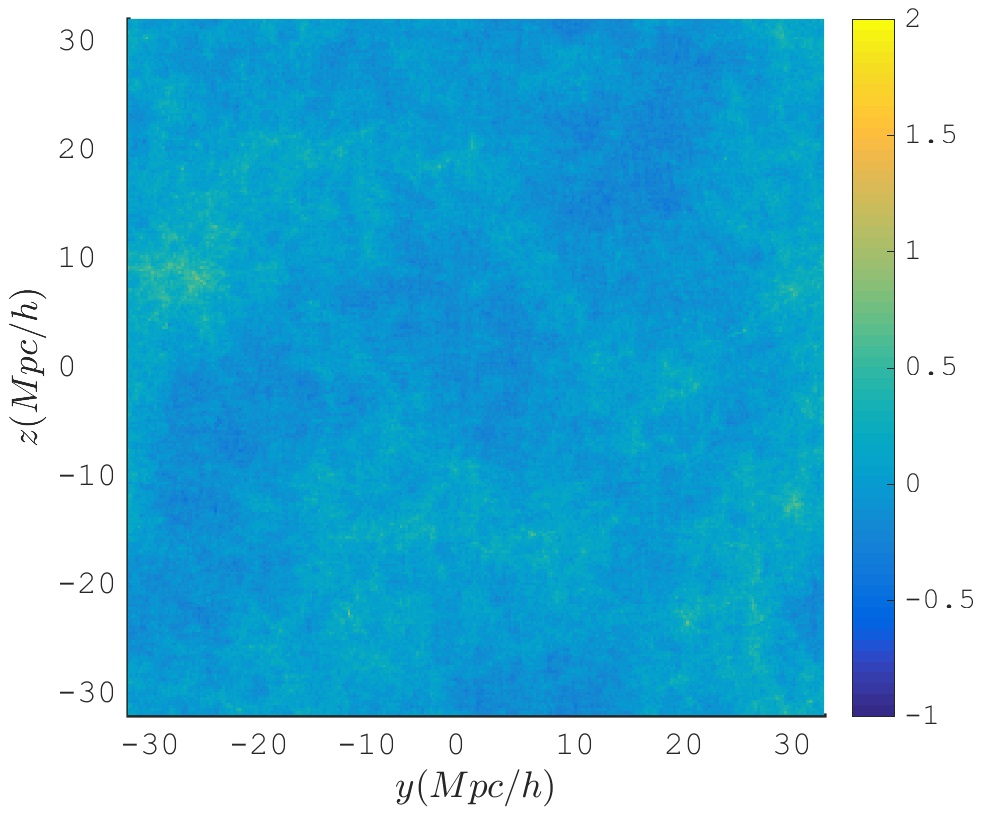}
	\caption{Density contrast of a 2D projection of the box onto the y-z plane at redshift $z = 10$ \cite{Disrael}.  The $64 \, h^{-1} {\rm{Mpc}} \, \times \, 64 \, h^{-1} {\rm{Mpc}} \, \times \,  64 \, h^{-1} {\rm{Mpc}}$ data box contains a wake caused by a cosmic string of tension $G\mu = 10^{-7}$ at the position y = 0 $h^{-1} {\rm{Mpc}}$.  The color scheme, shown on the right, shows the range of possible values for the fluctuation ratio $\delta \rho/\rho$ ($\rho$ being the density). The wake is not visible by eye.}
	\label{fig:wake_signal2}
\end{figure}

\begin{figure}[h]
	\centering
	\includegraphics[width = 7.5cm]{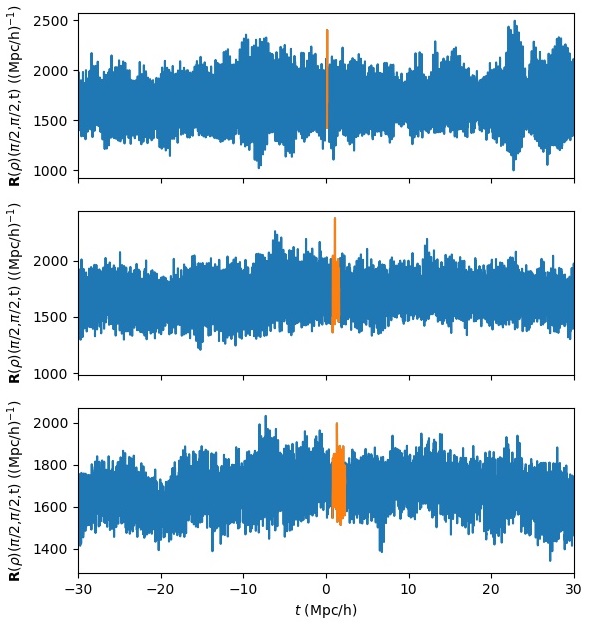}
	\caption{For a wake caused by a cosmic string of string tension $G\mu = 10^{-7}$ intruduced at redshift $z = 31$ at the center of the box in the x-z plane, the Radon transform $\textbf{R}(\rho) = \textbf{R}(\rho)(\theta_{1},\theta_{2},t)$ where $\theta_{1} = \theta_{2} =\pi/2$, is plotted for 10000 values of $t$ between -32 $h^{-1} {\rm{Mpc}}$ and 32 $h^{-1} {\rm{Mpc}}$. For this specific orientation and position, $t$ corresponds to the $y$ axis which gives us a density projection onto the y-axis.  The orange region shows how points in the original wake spread in the density projection at lower redshifts.}
	\label{fig:waketracking}
\end{figure}

At redshift $z = 10$, the typical 2D density contrast of a data box where a cosmic string wake was introduced shows no wake signal which is observable by eye.  This can be seen in Figure \ref{fig:wake_signal2}. 
Even though the wake signal cannot be observed in the 2D density projections, it can be extracted using ridgelet statistics.  In contrast to the example in the previous section where we study the wake at the time of insertion, we now study it at a later time. The wake slightly moves in position along the y-axis between the time when it is inserted ($z = 31$) and the time at which it is studied ($z = 10$).  In order to track the position of the wake between different redshifts, we index the ID of each particle in the overdensity made by the wake at $z = 31$.  Then, we let the system evolve to lower redshift while highlighting the interval in which the points in the initial overdensity are situated.  This process is shown in Figure \ref{fig:waketracking}. 
For the set of 4 simulations where the effects of a cosmic string are introduced, the mean value of the wake position at $z = 10$ is $(0.1 \pm 0.8) h^{-1} {\rm{Mpc}}$ which is consistent with the fact that the wake is inserted in the middle of the box at $y = 0 \, h^{-1}{\rm{Mpc}}$.

\begin{figure}[h]
	\centering
	\includegraphics[width = 7.5cm]{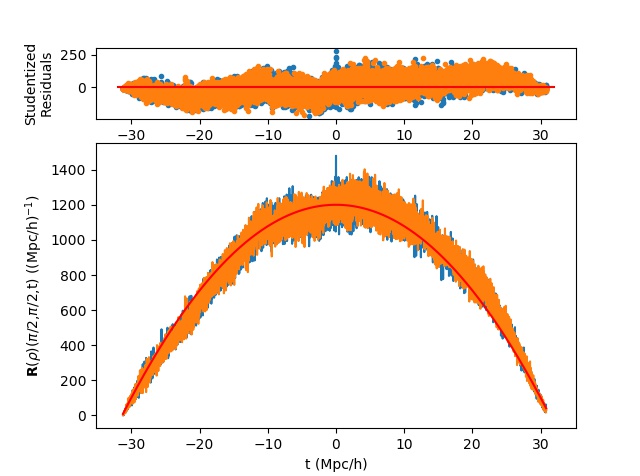}
	\caption{For a wake caused by a cosmic string of string tension $G\mu = 10^{-7}$ at redshift $z = 10$, the Radon transform of the data box with a wake (in blue) and without a wake (in orange) is plotted for 10000 values of t between $-32 \, h^{-1} {\rm{Mpc}}$ and $32 \, h^{-1} {\rm{Mpc}}$ for the fixed orientation $\theta_{1} = \theta_{2} = \pi/2$.  The Radon transform is fitted to as a second degree polynomial shown in red.  The studentized residuals reveal the presence of a small region of overdensity at $t = 0 \, h^{-1}{\rm{Mpc}}$ in the blue plot which is not found in the orange plot.  We conclude that this region of overdensity is caused by the wake.}
	\label{fig:radoninsphere}
\end{figure}

Once the wake is localized, the rest of the analysis is done on the points inside a sphere of radius of approximately $30 \, h^{-1} {\rm{Mpc}}$ centered on the wake.  To make sure that the wake was well centered in the sphere, we plotted the Radon transformation of this sphere for the fixed orientation $\theta_{1} = \theta_{2} = \pi/2$ in order to obtain the density projection of the sphere on the y-axis.  Then, the density projection on the y-axis was fitted to a second-degree polynomial. As shown in Figure \ref{fig:radoninsphere}, the studentized residuals of the fit exposes a peak at the center which confirms the presence of the wake at the center of the sphere for one of the simulations that were studied.
For the set of 4 simulations where a wake was introduced, the average measured width of the region of the peak was $(4.8 \pm 0.7) \, \times \, 10^{-2} h^{-1} {\rm{Mpc}}$.  In comparison, the same density projection for the box without the wake has no region of overdensity which confirms the presence of the wake in the data box and its location.

\begin{figure}[h]
	\centering
	\includegraphics[width = 7.0cm]{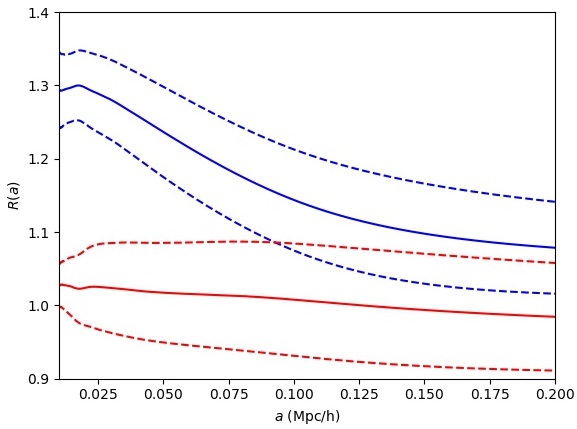}
	\caption{For a wake caused by a cosmic string of string tension $G\mu = 10^{-7}$ at redshift $z = 10$, the mean value of the ratio $R(a)$ is plotted for 3000 values of $a$ between $0.01 \, h^{-1} {\rm{Mpc}}$ and $0.20 \, h^{-1} {\rm{Mpc}}$.  The dashed lines show the standard deviation with respect to the means.  The mean ratio for the data boxes is plotted in blue while the mean ratios for the data boxes without wakes are plotted in red.}
	\label{fig:Scale_Ratios_2}
\end{figure}

In order to find the scale $a_{max}$ corresponding to the wakes, we fixed the position parameter $b$ at the center of the sphere $(b = 0$ $h^{-1} {\rm{Mpc}})$ for $\theta_{1} = \theta_{2} = \pi/2$ and plotted the ratio $R(a)$ for the set of data boxes with a wake and the set of boxes without the wake.  The mean value of the plots for the set of data boxes with a wake and the set of data boxes without a wake is shown in Figure \ref{fig:Scale_Ratios_2}.
The ratios for the data box with the wake show a maximum at $a_{max} = (1.7 \, \pm \, 0.1) \, \times \, 10^{-2}$ $h^{-1} {\rm{Mpc}}$.  Conversely, the ratios for the data without the wake (in red) stay close to 1.  Therefore, we conclude that $a_{max} = 1.7 \times 10^{-2}$ $h^{-1} {\rm{Mpc}}$ is a good fixed parameter for the scale parameter $a$.  Also, Figure \ref{fig:Scale_Ratios_2} gives us information on the range of scale parameters $a$ which allow the wake to be detected when compared to a set of simulations without a wake.  Indeed, if the lower bound on the mean of $R(a)$ for the data boxes with a wake is higher than the upper bound on the mean of $R(a)$ for the data boxes without a wake, we expect to be able to detect the wake for the given value of $a$.  In this case, we would be able to detect the wake for a value of $a$ lower than $0.90$ $h^{-1} {\rm{Mpc}}$, which is where the bounds of the two curves meet in Figure \ref{fig:Scale_Ratios_2}. 

\begin{figure*}[tb]
	\centering
	\includegraphics[width = 17.0cm]{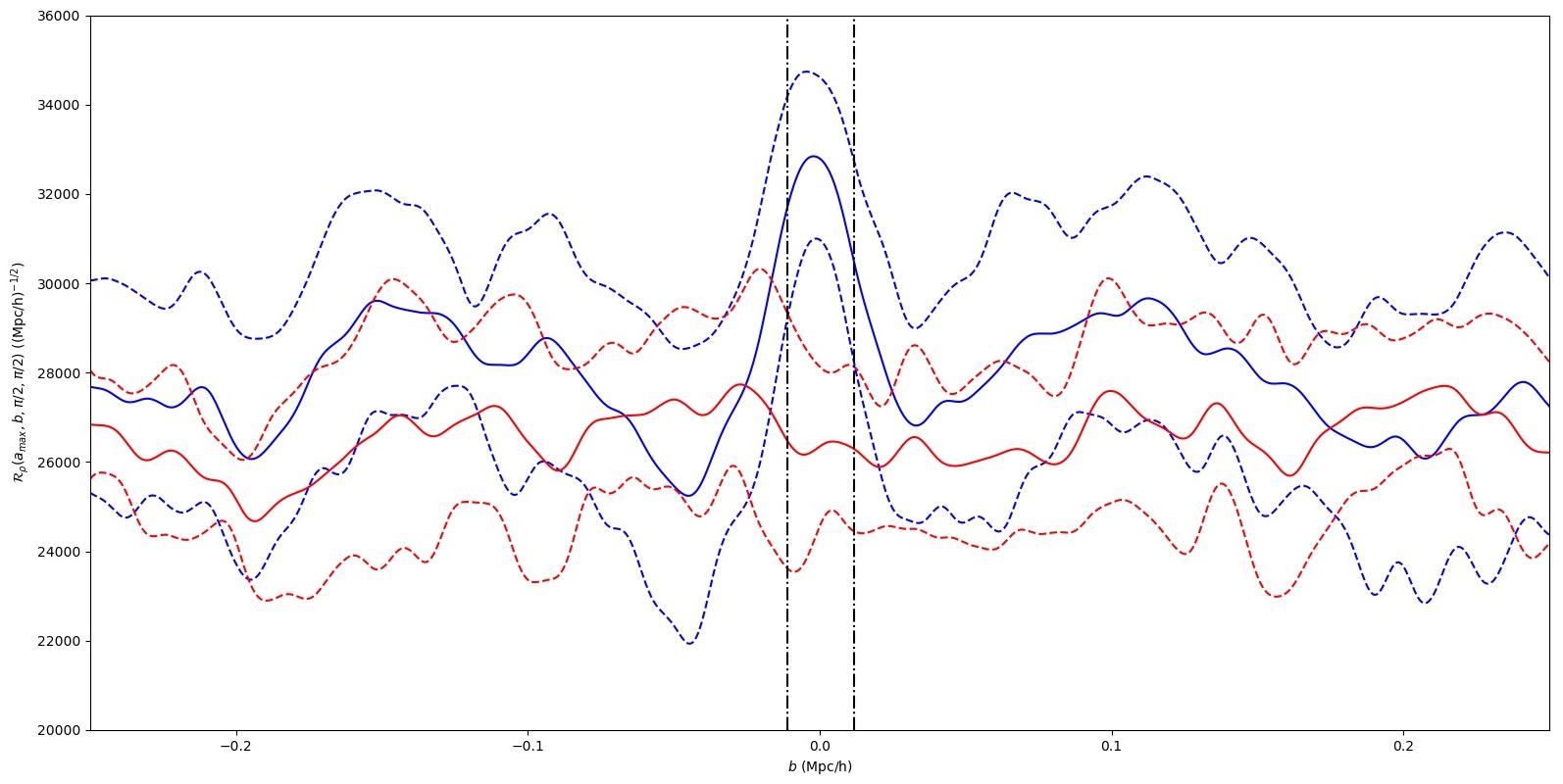}
	\caption{For a set of data boxes with a wake caused by a cosmic string of string tension $G\mu = 10^{-7}$ at redshift $z = 10$ (in blue) and a set of data boxes without wake (in red), the mean ridgelet coefficients $\mathcal{R}_{\rho}$ are plotted for 1000 values of $b$ between $-0.25 \,h^{-1}{\rm{Mpc}}$ and $0.25 \, h^{-1} {\rm{Mpc}}$ while imposing that $a = a_{max}$, $\theta_{1} = \theta_{2} =\pi/2$.  The dashed lines show the standard deviation with respect to the means and the vertical black lines show the interval in which the peak caused by the wake can be detected when compared to a set of simulations without wakes.}
	\label{fig:bplot2}
\end{figure*}

To ensure that $b = 0$ $h^{-1} {\rm{Mpc}}$ is a good parameter to impose for the Ridgelet transformation, we plotted $\mathcal{R}_{\rho}$ for $a = a_{max}$ and $\theta_{1} = \theta_{2} = \pi/2$ and different values of $b$.  The mean value of $\mathcal{R}_{\rho}$ for the set of data boxes with a wake and the set of data boxes without a wake is shown in Figure \ref{fig:bplot2}.
On average, the ridgelet coefficients have a maximum at $b_{max} = (-4 \, \pm \, 4) \, \times \, 10^{-3}$ $h^{-1} {\rm{Mpc}}$.  This agrees with the fact that $b = 0$ $h^{-1} {\rm{Mpc}}$ should maximize the ridgelet coefficients.  Therefore, $b = 0$ $h^{-1} {\rm{Mpc}}$ is a good fixed parameter.  In the same way as Figure \ref{fig:Scale_Ratios_2}, Figure \ref{fig:bplot2} gives us information on the range of position parameters $b$ which allow the wake to be detected when compared to a set of simulations without a wake.  This range of values,  marked by dashed black lines in the figure, goes from $-0.01 \, h^{-1} {\rm{Mpc}}$ to $0.01 \, h^{-1} {\rm{Mpc}}$.

\begin{figure*}[tb]
	\centering
	\includegraphics[width = 18.0cm]{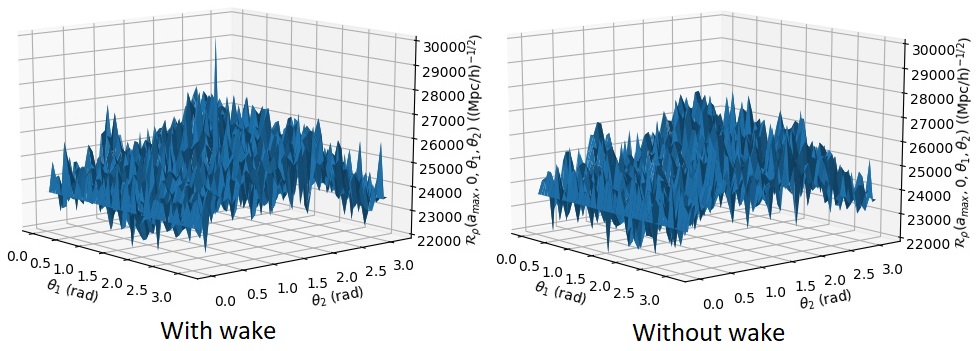}
	\caption{Surface plot of $\mathcal{R}_{\rho}$ for the data box with the wake caused by a cosmic string of string tension $G\mu = 10^{-7}$ at redshift $z = 10$ (on the left) and without a wake at the same redshift (on the right).  In both cases, the values of the ridgelet coefficients are computed for $a  = a_{max}$, $b = 0$ $h^{-1} {\rm{Mpc}}$ and $50^{2}$ orientations $(\theta_{1}, \theta_{2}) \in [0,\pi[ \, \times \,  [0,\pi[$.  For this example of data box with a wake, the coefficients have a clear maximum at $\theta_{1} = \theta_{2} =\pi/2$. This maximum is not observable in the coefficients of the data box without a wake.  Therefore, we conclude that the maximum in the ridgelet coefficients on the left plot is indeed caused by the wake.}
	\label{fig:Surface_Plot_2}
\end{figure*}

Finally, choosing $a = a_{max}$ and $b = 0$ $h^{-1} {\rm{Mpc}}$ as fixed parameters, we computed the partial ridgelet transformation for each data box with a wake and each data box without a wake.  The result is shown in Figure \ref{fig:Surface_Plot_2} for one round of simulations with and without wakes.  For each simulation, the ridgelet coefficients of the data box with the wake have a maximum at $\theta_{1} = \theta_{2} =\pi/2$.  On average, the confidence level was $\mathcal{C} = (6.8 \, \pm \, 0.9) \, \sigma$ for the simulations with a wake.  This confirms that the weak signal observed was indeed created by the cosmic string wake.

%%RB: This analysis needs to be improved. Something like peak to mean statistic. => SL: I changed the analysis to use the same definition as Disrael.

%%%%%%%%%%%%%%%
%% Conlusion %%
%%%%%%%%%%%%%%%

\section{Conclusions and Discussion}

We have applied 3D ridgelet transformations to output data from cosmological N-body simulations in which the effects of a cosmic string wake have been added to the standard $\Lambda$CDM fluctuations. The goal of our study was to determine down to which redshifts the wake signals remain visible for a string tension of $G \mu = 10^{-7}$, a tension slightly lower than the current upper bound. Given the limited resolution of our simulations, we found that the string signals can be extracted down to a redshift of $z = 10$. These results were reached by comparing the output data of N-body simulations with and without the effects of the string wake on the positions and velocities of the dark matter particles in the simulations. We expect that with higher resolution simulations the string signals remain visible to slightly lower redshifts. Our results are based on statistical analyses of a two-dimensional subspace of ridgelet coefficients, the other two having been fixed by independent considerations. An analysis of the full four-dimensional space of ridgelet coefficients will likely yield stronger limits. 

Our simulations yield the distribution of the dark matter. This could be probed observationally using weak lensing surveys. In order to compare with galaxy or quasar redshift surveys, our simulations would have to be extended with a halo-finding algorithm. On the other hand, we have seen that the wake signals rapidly get swamped by the non-Gaussianities due to the $\Lambda$CDM fluctuations. Hence, it may be more promising to study cosmic string signals in 21cm surveys at redshifts approaching that of reionization. 

\section*{Acknowledgments}

Two of us (RB and DC) are grateful to Adam Amara and Alexandre Refregier for discussions and encouragement. One of us (DC) thanks Joaquim Harnois-Deraps for help with the N-body code. This research has been supported in part by an NSERC Discovery Grant and by funds for the Canada Research Chair program. DC wishes to acknowledge CAPES (Science Without Borders) for a student
fellowship. This research was enabled in part by support provided by Calcul Quebec
http://www.calculquebec.ca/en/) and Compute Canada (www.computecanada.ca).

%%%%%%%%%%%%%%%%%%
%% Bibliography %%
%%%%%%%%%%%%%%%%%%

\end{document}